\baselineskip=18pt
\def\a{\alpha}\def\c{\chi}\def\d{\delta}
\def\f{\phi}\def\h{\theta}
\def\l{\lambda}\def\m{\mu}\def\n{\nu}\def
\p{\pi}\def\r{\rho}\def\s{\sigma}
\def\y{\eta}\def\x{\xi}\def\z{\zeta}

\def\D{\Delta}\def\L{\Lambda}
\def\O{\Omega}

\def\de{\partial}\def\na{\nabla}
\def\inf{\infty}\def\id{\equiv}\def\mo{{-1}}\def\ha{{1\over 2}}

\def\ds{ds^2=}

\def\af{asymptotically flat }
\def\fe{field equations }\def\bh{black hole }
\def\coo{coordinates }

\def\ssy{spherically symmetric }

\def\dof{degrees of freedom }

\def\sch{Schwarzschild }\def\ads{anti-de Sitter }
\def\RN{Reissner-Nordstr\"om }

\def\GB{Gauss-Bonnet }

\def\dsy{dynamical system }
\def\ab{asymptotic behaviour }

\def\section#1{\bigskip\noindent{\bf#1}\smallskip}
\def\nota{\footnote{$^\dagger$}}

\def\PL#1{Phys.\ Lett.\ {\bf#1}}
\def\PRL#1{Phys.\ Rev.\ Lett.\ {\bf#1}}
\def\PR#1{Phys.\ Rev.\ {\bf#1}}\def\CQG#1{Class.\ Quantum Grav.\ {\bf#1}}
\def\NP#1{Nucl.\ Phys.\ {\bf#1}}

\def\MPL#1{Mod.\ Phys.\ Lett.\ {\bf #1}} 
\def\PRep#1{Phys.\ Rep.\ {\bf#1}}

\def\grq#1{{\tt gr-qc/#1}}

\def\ref#1{\medskip\everypar={\hangindent 2\parindent}#1}
\def\beginref{\begingroup
\bigskip
\centerline{\bf References}
\nobreak\noindent}

\def\er{{\cal R}}\def\es{{\cal S}}\def\ef{e^{-2\f}}

\def\eg{e^\m}\def\el{e^\l}\def\eml{e^{-\l}}
\def\M{{\rm M}}\def\eff{e^{2\f}}

{\nopagenumbers \line{\hfil December 2005}
\vskip80pt
\centerline{\bf Global properties of charged dilatonic Gauss-Bonnet black
holes}
\vskip40pt
\centerline{{\bf M. Melis}\footnote{$^\dagger$}{e-mail:
maurizio.melis@ca.infn.it} and
{\bf S. Mignemi}\footnote{$^\ddagger$}{e-mail:
smignemi@unica.it}\footnote{$^*$}{and INFN, Sezione di Cagliari}}
\vskip10pt
\centerline {Dipartimento di Matematica, Universit\`a di Cagliari}
\centerline{viale Merello 92, 09123 Cagliari, Italy}

\vskip100pt
\centerline{\bf Abstract}
\vskip10pt
{\noindent
We study the phase space of the \ssy solutions of Einstein-Maxwell-\GB
system nonminimally coupled to a scalar field and prove the existence of
solutions with unusual asymptotics in addition to \af ones. We also find
new dyonic solutions of dilatonic Einstein-Maxwell theory.}
\vskip100pt\
P.A.C.S. Numbers: 97.60.Lf 11.25.Mj
\vfil\eject}

\section{1. Introduction}
In a recent paper [1] we have studied the global properties of the
\ssy solutions of Einstein-\GB (GB) gravity nonminimally coupled to
a scalar field. We refer to that paper for a full discussion of the
relevance of GB gravity for physics and for a list of references.
We only recall its role in the low-energy field theory expansion of
string theory [2].

Our investigation showed that in four dimensions the only regular
\bh solutions of the system are asymptotically flat. It is interesting
to check if this property holds also for charged black holes.
In fact, this is not the case. Already in absence of GB coupling
some degenerate charged solutions of  dilatonic gravity
with unusual asymptotic behaviour are known [3].
Actually, analytical solution to dilatonic gravity with nontrivial Maxwell
field are well known in absence of GB coupling [4,5]. If only magnetic or
electric charges are present, the solutions display only one horizon and
are analogous to \sch black holes.
If instead both charges are nonvanishing, they possess
an inner and an outer horizon and the causal structure is similar to that
of the \RN solution. Asymptotically, they are flat except in the
degenerate cases cited above [3].

In the case of dilatonic Einstein-GB models, magnetically charged solutions
cannot be obtained in analytical form, but have been studied perturbatively [6]
and numerically [7,8]. Only \af solutions have been investigated in this
context.
The existence of solutions with different asymptotics can however be decided
through the study of the global properties of the phase space associated with
the dynamical system, as has been shown in the literature in analogous situations
[9].
It must be noticed that the search for solution is made easier in the
Einstein case by the existence of a duality between electrically and
magnetically charged solutions, that is no longer available in the presence
of GB corrections.

Also interesting is the study of the emergence of branch singularities in
the metrics. This is a common feature of GB models [10,7], and can give rise
to naked singularities. This topic can be investigated by means of an expansion
of the fields near the outer horizon [11].

In this paper, we study the dilatonic Einstein-GB system coupled to Maxwell field,
and classify the solutions by their \ab. We find that their behaviour is
the same as in the dilatonic Einstein case, except for some non-\af purely magnetic
solutions, which do not have analogues if GB terms are included.
In the course of the investigation, we also find new dyonic solutions of
dilatonic Einstein gravity with unusual asymptotics. We also study
the near-horizon expansion of the metric functions, finding the
conditions for the absence of naked branch singularities and show
the existence of solutions with multiple horizons.

\section{2. The dynamical system}
In this section, using the methods of ref.\ [1], we derive the dynamical system
associated to \ssy solutions of the field equations of dilatonic Einstein-GB-Maxwell
theory.
For further technical details we refer to that paper.

The four-dimensional action of four-dimensional Einstein-GB gravity coupled to
electromagnetism is
$$I={1\over16\p G}\int\sqrt{-g}\ d^4x
\left[\er-2(\de\f)^2+\a\ef(\es-F^2)\right],\eqno(1)$$
where $\es=\er^{\m\n\r\s}\er_{\m\n\r\s}-4\er^{\m\n}\er_{\m\n}+\er^2$
is the GB term, $F$ is the Maxwell field, $\f$ is a scalar field and $\a$ is a
constant with dimension (length)$^2$. This action also emerges in the low-energy
expansion of string theory [2], and in this case the constant $\a$ is
proportional to the slope parameter. Note, however, that in string theory also an
axion field may couple to the gravitational and Maxwell fields, and this may change
notably the properties of the solutions [12].

For the metric we adopt the \ssy ansatz
$$ds^2=-\D\, dt^2+\D^\mo\, dr^2+R^2d\O^2,\eqno(2)$$
where $\D$ and $R$ are functions of $r$.
In these coordinates the generalized Maxwell equations
$$\na^\m(e^{-2\f}F_{\m\n})=0\eqno(3)$$
admit the \ssy solution
$$F=q_e{e^{2\f}\over R^2}\ dt\wedge dr+q_m\sin\h\ d\h\wedge d\f,\eqno(4)$$
where $q_e$ and $q_m$ are the electric and the magnetic charge,
respectively.

The \fe for the gravitational and scalar \dof following from (1) and (4) are
more conveniently written in terms of new variables such that the metric and
the scalar take the form [1]
$$ds^2=-e^{\y+\c}dt^2+e^{4\z-\y-\c}d\x^2+e^{2\z-\y-\c}d\O^2,
\qquad2\f=\y-\c,\eqno(5)$$
where $\z$, $\y$, $\c$ and $\l$ are functions of $\x$. They read
$$\eqalignno{
&\z''-\a[(\y'^2-\c'^2)(\y'+\c'-2\z')e^{2\c-4\z}]'=e^{2\z}+\a(\y'^2-\c'^2)
[2e^{2\c-2\z}-(\y'+\c'-2\z')^2e^{2\c-4\z}],&\cr
&\y''+\a[4\y'e^{2\c-2\z}-(\y'+\c'-2\z')(2\y'^2-\c'^2+\y'\c'-2\y'\z')
e^{2\c-4\z}]'=2\a q_e^2e^{2\y},&\cr
&\c''-\a[4\c'e^{2\c-2\z}+(\y'+\c'-2\z')(\y'^2-2\c'^2-\y'\c'+2\y'\z')e^{2\c-4\z}]'=
&\cr
&\qquad\qquad\qquad\qquad\quad2\a q_m^2e^{2\c}+\a(\y'^2-\c'^2)[4e^{2\c-2\z}-
(\y'+\c'-2\z')^2e^{2\c-4\z}],&\cr
&2\z'^2-\y'^2-\c'^2-\a(\y'^2-\c'^2)[4e^{2\c-2\z}-3(\y'+\c'-2\z')^2e^{2\c-4\z}]=
2e^{2\z}-2\a q_m^2e^{2\c}-2\a q_e^2e^{2\y}.&(6)\cr}$$
It is interesting to notice that in spite of the
presence of the higher-derivative GB term, the \fe are second order.

We can then write (6) in the form of a first order dynamical system by
defining new variables
$$X=\z',\quad Y=\y',\quad W=\c',\quad Z=e^{\z},\quad U=\sqrt\a\,e^{\y},
\quad V=\sqrt\a\,e^{\c},\eqno(7)$$
which satisfy the differential equations
$$Z'=XZ,\qquad U'=YU,\qquad V'=WV.\eqno(8)$$
In terms of these variables, the \fe become
$$\eqalignno{
&X'-\left\{(Y^2-W^2)(Y+W-2X){V^2\over Z^4}\right\}'=
Z^2+(Y^2-W^2)\Big[2Z^2-(Y+W-2X)^2\Big]{V^2\over Z^4},&(9)\cr
&Y'+\left\{\Big[4YZ^2-(Y+W-2X)(2Y^2-W^2+YW-2XY)\Big]{V^2\over Z^4}
\right\}'=2q_e^2U^2,&(10)\cr
&W'-\left\{\Big[4WZ^2+(Y+W-2X)(Y^2-2W^2-YW+2XW)\Big]{V^2\over Z^4}\right\}'=&\cr
&\qquad\qquad\qquad\qquad\qquad\quad2q_m^2V^2+(Y^2-W^2)\Big[4Z^2-(Y+W-2X)^2\Big]
{V^2\over Z^4},&(11)}$$
subject to the constraint
$$E=2X^2-Y^2-W^2-2Z^2+2q_m^2V^2+2q_e^2U^2-(Y^2-W^2)
\Big[4Z^2-3(Y+W-2X)^2\Big]{V^2\over Z^4}=0.\eqno(12)
$$
In general, the \dsy is six-dimensional, but for purely magnetic
solutions ($q_e=0$), the variable $U$ is ignorable and does not appear in the
system, while (10) yields the first integral
$$Y+\Big[4YZ^2-(Y+W-2X)(2Y^2-W^2+YW-2XY)\Big]{V^2\over Z^4}=a.\eqno(13)$$
In this case the \dsy is effectively four-dimensional. We shall nevertheless
work also in this case with the full six-dimensional phase space, because this
facilitates the discussion of the critical points.
For the same reason, we shall not use the constraint (12) to eliminate $V^2/Z^4$
and reduce by one the dimensionality of the phase space.

The system can be put in canonical form solving (9)-(11) for the
first derivatives, as
$$X'=F(X,Y,W,Z,U,V),\qquad Y'=G(X,Y,W,Z,U,V),\qquad W'=H(X,Y,W,Z,U,V).\eqno(14)$$
The expressions so obtained are awkward and shall not be reported here.

\section{3. Charged solutions of dilatonic Einstein gravity}
In the following, the knowledge of the exact solutions
in absence of the GB term will prove useful. These were first discussed
in [4], and later some degenerate cases with nontrivial
asymptotics were found in [2] for the magnetic monopole. We shall
generalize these solutions to the dyonic case.
An important property of the dilatonic Einstein-Maxwell theory is
the presence of a duality invariance for the interchange of $q_e$
and $q_m$, with $\f\to-\f$.

In absence of \GB contributions, the \fe (6) reduce to
$$\z''=e^{2\z},\qquad\y''=2q_e^2e^{2\y},\qquad \c''=2q_m^2e^{2\c},\eqno(15)$$
subject to the constraint
$$2\z'^2-\y'^2-\c'^2-2e^{2\z}+2q_m^2e^{2\c}+2q_e^2e^{2\y}=0,\eqno(16)$$
where we have absorbed $\a$ into the definition of the charges. In these
variables, the duality invariance corresponds to the interchange of $\y$ and
$\c$.
The integration of (15) yields
$$e^\z={2a\,e^{a\x}\over1-e^{2a\x}},\qquad
e^\y={\sqrt2\over|q_e|}{b\,e^{b(\x-\x_1)}\over1-e^{2b(\x-\x_1)}},\qquad
e^\c={\sqrt2\over|q_m|}{c\,e^{c(\x-\x_2)}\over1-e^{2c(\x-\x_2)}},\eqno(17)$$
where $a$, $b$, $c$, $\x_1$ and $\x_2$ are integration constants
and the origin of $\x$ has been fixed arbitrarily.
Substituting in the constraint (16), one gets $2a^2-b^2-c^2=0$.
If one also requires that the solution has a regular horizon, one
must impose that $R^2=e^{2\z-\y-\c}$ is regular at the horizon, $\x\to-\inf$,
which implies $2a-b-c=0$. Combined with the previous constraint, this yields
$a=b=c$.
Substituting these results in (5) and defining a new
coordinate $r=\int e^{2\z}d\x=2a/(1-e^{2a\x})$, one obtains the solution
$$\eqalignno{ds^2&=-{r(r-r_h)\over(r+r_e)(r+r_m)}\ dt^2+{(r+r_e)(r+r_m)
\over r(r-r_h)}\ dr^2+(r+r_e)(r+r_m)\ d\O^2,&\cr
\ef&={(r+r_e)\over(r+r_m)},&(18)}$$
where $r_h=2a$ and $r_e$ and $r_m$ are functions of $a$, $\x_1$ and $\x_2$.
This solution was first discussed in ref.\ [4].
The mass $M$ and the charges of the black hole are related to the parameters
of the solution by
$$2M=r_h+r_e+r_m,\qquad q_e^2={r_e(r_h+r_e)\over2},\qquad q_m^2={r_m(r_h+r_m)
\over2}.\eqno(19)$$
The metric (18) has two horizons at $r=r_h$ and
$r=0$, a singularity at $r=\max(-r_e,-r_m)$, and is asymptotically flat. The
special case when $q_e=0$ was also derived in [5] and corresponds to $r_e\to0$.

Some degenerate solutions with unusual asymptotics arise when $\x_1$
or $\x_2$ vanish. For example, for $\x_2=0$, one has
$$ds^2=-{r(r-r_h)\over r+r_e}\ dt^2+{r+r_e\over r(r-r_h)}\ dr^2+
(r+r_e)\,d\O^2,\qquad\ef={r+r_e\over2q_m^2}.\eqno(20)$$
For $\x_1=0$, the
metric has the same form, with $r_e\to r_m$, while the scalar
becomes $\ef=2q_e^2/(r+r_m)$. In
different \coo the metric (20) can be also written
$$ds^2=-{(R^2-r_e)(R^2-r_0)\over R^2}\ dt^2+
{R^4\over(R^2-r_e)(R^2-r_0)}\ dR^2+R^2\,d\O^2,\eqno(21)$$
with $\ef=R^2/2q_m^2$, and $r_0=r_h+r_e$.
The solution (20) generalizes that found in [3] for the purely magnetic
case, which is recovered in the limit $r_e\to0$.
In the general case, the metric (20) displays an outer horizon
at $r=r_h$, an inner horizon at $r=0$, and a singularity at $r=-r_e$.
Its \ab is intermediate between \sch and \ads and its causal structure is
similar to that of the \RN metric.

For $\x_1=\x_2=0$, one obtains a previously unknown solution, of the form
$$ds^2=-{r(r-r_h)\over2q_eq_m}dt^2+{2q_eq_m\over r(r-r_h)}dr^2+2q_eq_md\O^2,
\qquad\ef={q_e\over q_m}.\eqno(22)$$
This metric is the direct product of two-dimensional \ads spacetime with
a two-dimensional sphere and is therefore of the Bertotti-Robinson form.
It has a horizons at $r=0$ and $r=r_h$, but no singularity, and both
nonvanishing electric and magnetic charge.

In all cases, extremal black holes are obtained in the limit $a=0$, corresponding
to $r_h=0$, or, in terms of the physical parameters, $2M^2=(q_e\pm q_m)^2$.

Finally, we notice that in terms of the variables of the dynamical system the
solution (17) reads
$$X=a\coth(a\,\x),\qquad Y=a\coth[a(\x-\x_1)],\qquad W=a\coth[a(\x-\x_2)],$$
$$Z={a\over\sinh(a\,\x)},\quad U={a\over\sqrt2|q_e|\sinh[a(\x-\x_1)]},
\quad V={a\over\sqrt2|q_m|\sinh[a(\x-\x_2)]},$$
\smallskip
{\noindent with the horizon at $\x\to-\inf$, and spatial infinity at $\x=0$.
From the explicit knowledge of the solutions or by standard
methods, it is then easy to construct the phase space portrait.
Phase space is invariant under $Z\to-Z$, $U\to-U$ or $V\to-V$,
thus we shall limit our study to positive values of these variables.}

The physical trajectories lie on the five-dimensional hyperplane $E=0$.
The critical point at finite distance lie on the surface
$Z_0=U_0=V_0=P_0=0$, where
$$P^2\id 2X^2-Y^2-W^2,\eqno(23)$$
but only points with $X_0=Y_0=W_0$ correspond to regular horizons. The
eigenvalues of the linearized equations around the critical points are
$0(3)$, $X_0$, $Y_0$, $W_0$.

The phase space at infinity can be studied defining new variables
$$t={1\over X},\quad y={Y\over X},\quad w={W\over X},\quad z={Z\over X},
\quad u={U\over X},\quad v={V\over X}.\eqno(24)$$
The critical points at infinity are found at $t_0=0$ and

a) $u_0=v_0=z_0=0$, $y=y_0$, $w=w_0$, with $y_0^2+w_0^2=2$. These
are endpoints of trajectories lying on the surface at infinity.

b) $u_0=v_0=0$, $z_0=1$, $y_0=w_0=0$. These are the endpoints
of the \af solutions (18).

c) $u_0=0$, $v_0=1/\sqrt{2q_m^2}$, $z_0=1$, $y_0=0$, $w_0=1$.
These are the endpoints of the solutions (20).

d) $v_0=0$, $u_0=1/\sqrt{2q_e^2}$, $z_0=1$, $y_0=1$, $w_0=0$.
These are the endpoints of the dual solutions to (20).

e) $u_0=1/\sqrt{2q_e^2}$, $v_0=1/\sqrt{2q_m^2}$, $z_0=1$,
$y_0=w_0=1$. These are the endpoints of the solutions (22).

{\noindent The eigenvalues of the linearized equations are}

a) 0(3), 1, $y_0$, $w_0$.

b) $-1(5)$, $-2$.

c) $1$, $-1(3)$, $-2(2)$.

d) $1$, $-1(3)$, $-2(2)$.

e) $1(2)$, $-1$, $-2(3)$.

{\noindent In absence of electric charge, the system reduces to a
four-dimensional system with $U=Y=0$, and similarly for $q_m=0$,
it reduces to a system with $V=W=0$.}

The phase space portrait is the following: regular solutions start at
$Z_0=U_0=V_0=0,\ $ $X_0=Y_0=W_0=a$, for some value of the parameter $a$.
At infinity, a three-dimensional family of trajectories is attracted by points b)
and a two-dimensional bunch by points c) and d). Finally, a one-dimensional bunch
ends at the critical point e).

\section{4. Phase space of the Einstein-GB system}

We discuss now the phase space associated with the Einstein-GB system (8), (14).
Since $E$ in (12) is a first integral of the dynamical system, all the
trajectories are confined to surfaces of constant $E$, but only those with
vanishing $E$ are physical. We shall therefore consider only the critical
points lying on the hypersurface $E=0$. The discussion of the phase space is
complicated by the presence of a singularity in the field
equations at $Z=0$. Since some of the critical points of the \dsy lie on this
hypersurface, one must be careful in taking the limits when approaching these
points. In particular, when $Z\to0$, also $V\to0$.
Although the duality invariance is no longer valid, the phase space is still
invariant under $Z\to-Z$, $U\to-U$ or $V\to-V$, as in the pure Einstein case.
We can therefore consider only positive values of $Z$, $U$ and $V$.

The critical points at finite distance are defined as the points were the r.h.s.\
of eqs.\ (8) and (14) vanishes, and in analogy with the Einstein case are found to
lie on the surface $Z_0=U_0=V_0=P_0=0$, with $P$ given by (23).
We are interested in regular \bh solutions. A regular horizon is present
if $R^2\id e^{2\z-\y-\c}$ is finite at the critical points. Moreover,
the dilaton is regular at the horizon if $e^{-2\f}\id e^{\y-\c}$ is
finite there. These conditions are equivalent to the requirement that at
the horizon $Z^2\sim U^2\sim V^2$. Since the critical points correspond to
$\x\to\pm\inf$, with $Z^2\sim e^{2X_0\x}$, $U^2\sim e^{2Y_0\x}$,
$V^2\sim e^{2W_0\x}$, one obtains that the horizon is regular iff
$X_0=Y_0=W_0=a$, for some real parameter $a$.
To simplify the following discussion, we shall only consider points that
satisfy this constraint.

The nature of the critical points can be investigated linearizing (8) and (14)
around them. One obtains
$$\d Y'=\d X'=\d W'=0,\quad\d Z'=a\,\d Z,\quad\d U'=a\,\d U,\quad\d V'=a\,\d V,$$
with eigenvalues $0$ (3), $a$ (3). It results therefore that the critical point
is an attractor for $a<0$ and a repellor for $a>0$.

The \ab of the solutions is dictated by the nature of the critical
points at infinity [8,1]. These can be investigated using the variables (24).
The \fe at infinity ($t\to0$) read
$$\eqalign{
&\dot t=-ft,\qquad\qquad\dot y=g-fy,\qquad\ \dot w=h-fw,\cr &\dot
z=(1-f)z,\qquad\dot u=(y-f)u\qquad\dot v=(w-f)v,}$$
where a dot denotes $t\,d/d\x$, $f=t^2F(1,y,w,z,u,v)$, $g=t^2G(1,y,w,z,u,v)$,
$h=t^2H(1,y,w,z,u,v)$, and the variables are subject to the constraint
$$2-y^2-w^2-2z^2+2q_m^2v^2+2q_e^2u^2-(y^2-w^2)
\Big[4z^2-3(y+w-2)^2\Big]{v^2\over z^4}=0.$$

The critical points at infinity are placed at $t_0=0$, with

a) $u_0=v_0=z_0=0$, $y=y_0$, $w=w_0$, with $y_0^2+w_0^2=2$.

b) $u_0=v_0=0$, $z_0=1$, $y_0=w_0=0$.

c) $v_0=0$, $u_0=1/\sqrt{2q_e^2}$, $z_0=1$, $y_0=1$, $w_0=0$.

e) $u_0=\sqrt{q_m^2+4\over(2q_m^2+4)q_e^2}$,
$v_0={1\over\sqrt{2q_m^2+4}}$, $z_0=1$, $y_0=w_0=1$.
\medskip
{\noindent The points a), b) and c) coincide with those obtained in the
Einstein case, while the point e) is displaced due to the GB contribution. The most
interesting feature is however that the critical point d) has disappeared.
In particular, no solution with vanishing electric charge and unusual asymptotics
exists in the GB case.

The eigenvalues of the linearized equations are}

a) 0(3), 1, $y_0$, $w_0$.

b) $-1(5)$, $-2$.

c) $1$, $-1(3)$, $-2(2)$.

e) $1$, $-1$, $-2(2)$, $-\ha(1\pm\sqrt{9+16/q_m^2})$.

{\noindent They have the same sign as in the Einstein case, and hence analogous
properties.}

The resulting phase space portrait is rather similar to that of the Einstein
system, except for the critical points d), that in the GB case are absent.
In particular, all the trajectories corresponding to solutions with regular
horizons start from the critical point at $Z_0=U_0=V_0=0$, $X_0=Y_0=W_0=a>0$
for a given value of the parameter $a$, and end at the points b), c) or e).

From the location of the critical points at infinity, one can
deduce the asymptotic behaviour of the solutions [9,1]. The behaviour
coincides in all cases with that found in the Einstein limit. In particular,
in case b)
solutions are \af with asymptotically constant scalar field, while in case c)
they behave like (20) and in case e) like (22). The existence of solutions
with $a=0$ also suggests the existence of metrics with multiple horizons,
i.e.\ with multiple roots of the metric functions, that should correspond to
extremal black holes.

\section{5. Branch singularities}
In the previous section, we have studied the existence of solutions with regular
horizons. Our method, however, does not give a complete characterization of the
black holes. In particular, it does not permit to study the region inside the
horizon and to ascertain the existence of inner horizons\nota{Some results on
this topic have been discussed using numerical methods in [7].}. Moreover, we are
not able to relate the parameters $a$, $q_m$ and $q_e$ to the radius of the horizon.
On the other hand, in GB models, branch singularities usually appear [10,7]:
the requirement that they are shielded by an horizon imposes a minimal value for
the horizon radius (and consequently for the mass of the black hole). In order to
investigate the presence of these singularities, it is useful to study the
behaviour of the fields near the outer horizon [11].
The calculation is most easily performed in coordinates of the form
$$\ds-e^{\m(R)}dt^2+e^{\l(R)}dR^2+R^2d\O^2.\eqno(25)$$

In this gauge, the \fe read
$$\eqalignno{
&\f''+\left({\m'-\l'\over2}+{2\over R}\right)\f'=&\cr
&\qquad-{2\a\ef\over R^2}
\left[\m'\l'\eml+(1-\eml)\left(\m''+{\m'\over2}(\m'-\l')\right)\right]
-\a(q_m^2\ef-q_e^2\eff){\el\over R^4},&(26)\cr
&{\l'\over R}-{1-e^\l\over R^2}=\f'^2+{4\a\over R^2}\ef\left[(1-3\eml)\m'\f'-
2(1-\eml)(\f''-2\f'^2)\right]+\a(q_m^2\ef+q_e^2\eff){\el\over R^4},&(27)\cr
&{\m'\over R}+{1-e^\l\over R^2}=\f'^2+{4\a\over R^2}\ef(1-3\eml)\m'\f'
-\a(q_m^2\ef+q_e^2\eff){\el\over R^4},&(28)\cr
&\m''+\left({\m'\over2}+{1\over R}\right)(\m'-\l')=&\cr
&\qquad-2\f'^2-{8\a\ef\over R}\eml\left[\left(\m''+{\m'\over2}
(\m'-3\l')\right)\f'+\m'(\f''-2\f'^2)\right]
+2\a(q_m^2\ef+q_e^2\eff){\el\over R^4}.&(29)}$$

In order to find an analytical expression for the minimal value of the radius,
we assume that near the horizon $R=R_h$ the fields behave as
$$\eml=\L(R-R_h)+\dots,\qquad \eg=\M(R-R_h)+\dots,$$
$$\f=\f_h+\f'_h(R-R_h)+\dots,$$
for some constants $\L$, $\M$, $\f_h$ and $\f'_h$.
Let us first consider the case $q_e=0$.
Substituting into the \fe, one obtains from (28)
$$\L={1\over R_h}{1-g_m\over1-f},\eqno(30)$$
where
$$g_m={\a q_m^2e^{-2\f_h}\over R_h^2},\qquad f={4\a\f'_he^{-2\f_h}\over R_h}.
\eqno(31)$$

Moreover, combining (26) and (29), it results that
$$f={8\a^2e^{-4\f_h}\over R_h^3}\left[\L+{2\over R_h}\left(1+\L R_h f-
{g_m\over\L R_h}\right)\right]-{4\a e^{-2\f_h}g_m\over\L R_h^3},\eqno(32)$$
and hence, substituting (30) in (32), after some algebraic manipulations one
obtains a second degree algebraic equation for $f$:
$$[1-(1+\bar\a+\bar\a^2)g_m]f^2-[1-(1+2\bar\a+\bar\a^2)g_m+\bar\a^2g_m^2]f
+\ha[3\bar\a^2-2\bar\a(1+3\bar\a)g_m+\bar\a^2g_m^2]=0,\eqno(33)$$
where we have defined $\bar\a=4\a e^{-2\f_h}/R_h^2$.

In order to have real solutions for $f$, the discriminant $\D$ of the equation
must be nonnegative, i.e.
$$\D=(1-g_m)^2[1-6\bar\a^2+2\bar\a^2(1+3\bar\a+3\bar\a^2)g_m+\bar\a^4g_m^2]\ge0.
\eqno(34)$$
The inequality holds for any $g_m$ as long as $\bar\a^2\le1/6$, i.e.
$R_h^2\ge\a e^{-2\f_h}/4\sqrt6$. The same restriction holds also for neutral
solutions [11,7]. However, in the charged case (34) can be satisfied also when
$\bar\a^2\ge1/6$, if
$$g_m\ge{-(2+2\bar\a+3\bar\a^2)+\sqrt{3(1+4\bar\a+9\bar\a^2+6\bar\a^3+3\bar\a^4)}
\over\bar\a^2}.$$
The same result was also obtained in [8] for SU(2) Yang-Mills monopoles.
It is also interesting to remark that the vanishing of the prefactor in (34),
$g_m=1$, corresponds to extremal black holes.

When $q_e\ne0$, one obtains similar results. In this case,
defining $g_e=\a e^{2\f_h}\,q_e^2/R_h^2$,
$$\D=(1-g_m-g_e)^2[1-6\bar\a^2+2\bar\a^2(1+3\bar\a^2)(g_m+g_e)+
\bar\a^4(g_m+g_e)^2+6(g_m-g_e)],\eqno(35)$$
which is nonnegative for $\bar\a^2\le1/6$. If instead $\bar\a^2\ge1/6$ a
complicated constraint must be obeyed by the charges. Again, the case
$g_m+g_e=1$ corresponds to the extremal limit.

\section{6. Conclusions}
The study of charged solutions of the dilatonic Einstein-GB model
has revealed the presence of some difference from the pure Einstein case.
In particular, non-\af solutions with vanishing electric charge are absent
in the GB case.
The conditions for the absence of naked branch singularity is
identical to that of the neutral case, but the possibility exists
of solutions violating this condition, if the charges satisfy
some inequalities.
Also interesting is the existence of extremal solutions, which
were not considered in previous investigations.

\beginref
\ref [1] M. Melis and S. Mignemi, \CQG{22}, 3169 (2005).
\ref [2]  B. Zwiebach, \PL{B156}, 315 (1985);
B. Zumino, \PRep{137}, 109 (1986);
D.J. Gross and J.H. Sloan, \NP{B291}, 41 (1987).
\ref [3] K.C.K. Chan, J.H. Horne and R. Mann, \NP{B 447}, 441 (1995);
\ref [4] G.W. Gibbons and K. Maeda, \NP{B298}, 741 (1988).
\ref [5] D. Garfinkle, G.T. Horowitz and A. Strominger, \PR{D43}, 3140 (1991).
\ref [6] S. Mignemi and N.R. Stewart, \PR{D47}, 5259 (1993);
S. Mignemi, \PR{D51}, 934 (1995).
\ref [7] T. Torii, H. Yajima, and K. Maeda, \PR{D55}, 739 (1997);
S.O. Alexeyev and M.V. Pomazanov, \grq{9706066}.
\ref [8] P. Kanti and K. Tamvakis, \PL{B392}, 30 (1997).
\ref [9] S. Mignemi and D.L. Wiltshire, \CQG{6}, 987 (1989);
D.L. Wiltshire, \PR{D44}, 1100 (1991);
S. Mignemi and D.L. Wiltshire, \PR{D46}, 1475 (1992);
S.J. Poletti and D.L. Wiltshire, \PR{D50}, 7260 (1994);
S. Mignemi, \PR{D62}, 024014 (2000).
\ref [10] D.G. Boulware and S. Deser, \PRL{55}, 2656 (1985);
D.L. Wiltshire, \PL{B169}, 36 (1986).
\ref [11] P. Kanti,  N.E. Mavromatos, J. Rizos,  K. Tamvakis and E. Winstanley,
\PR{D54}, 5049 (1996).
\ref [12] A. Shapere, S. Trivedi and F. Wilczek, \MPL{A6}, 2677 (1991).
\end